\documentclass[twocolumn,showpacs,preprintnumbers,prl,aps,apssymb]{revtex4-1}
\usepackage{amsmath,amssymb}
\usepackage{graphicx}
\usepackage{dcolumn}
\usepackage{bm}
\usepackage{psfrag}
\usepackage{subfigure}
\newcommand{\ber}{\begin{eqnarray}}
\newcommand{\eer}{\end{eqnarray}}
\newcommand{\bea}{\begin{equation}}
\newcommand{\eea}{\end{equation}}

\begin{document}

\title{Revisiting Langer-Ambegaokar-McCumber-Halperin theory of resistive transitions in one-dimensional superconductors with exact solutions.}

\author{Darshan G. Joshi and A. Bhattacharyay}
\email{a.bhattacharyay@iiserpune.ac.in}
\affiliation{Indian Institute of Science Education and Research, Pune, India}

\date{\today}

\begin{abstract}
We present an important correction to the Langer-Ambegaokar-McCumber-Halperin theory for the resistive state of a 1D superconductor. We establish that the identification of the saddle on the free energy surface over which Langer and Ambegaokar had claimed the system to move in order to form thermally excited phase slip centres is wrong. With the help of an exact solution we show that the system has to overcome a similar free energy barrier but can actually have vanishing amplitude of superconducting phase at a point unlike the Langer-Ambegaokar solution.
\end{abstract}
\maketitle
In a 1D superconductor, kept below the critical temperature $T_C$, a current driven transition to a normal (N) phase from the superconducting (SC) phase has been seen to occur in experiments as early as in 1967 \cite{parks,hunt}. Such transitions show a finite width in temperature and current strength within which SC and N phases coexist. A theory was proposed by Langer and Ambegaokar (LA) \cite{lan}, which was subsequently improved by McCumber and Halperin (MH) \cite{mac}, to explain thermal fluctuations induced transitions through metastable states. Th Langer-Ambegaokar-McCumber-Halperin (LAMH) theory traces the origin of localized N phases in superconducting 1D (width is smaller than the coherence length $\xi$) samples in the formation of phase slip centres (PSC) - an idea which was originally put forward by Little \cite{lit}. Over last 40 years, the LAMH theory successfully accounts for the resistivity vs temperature plots observed in this (so-called) resistive regime except for some stretches at the lower temperature end of the resistivity vs temperature plot. At this end (near the $T_C$), where the resistivity vanishes and the system moves into the SC phase, some mismatch with the theoretically predicted values have been observed in the early experiments which were believed to be effects of imperfections at the contacts at the ends of the sample \cite{tink}. These deviations have also been seen to be system specific particularly lending support to relating their origin to contact imperfections. Classic experiments done by Webb-Warburton \cite{web} and Newbower {\it et al} \cite{new} showed the excellent applicability of LAMH theory to experimental results. Since then, the LAMH theory has remained the basic tool for dealing with the resistive regime of the 1D superconductor near the $T_C$ regime where quantum effects are negligible.
\par
Such a successful theory, however, has a particular lacuna which is a glaring inconsistency in it. The SC order parameter is a complex number and in 1D it looks like a spiral wound around the wire. The (LAMH) theory is based on a suggestion by Little, that, the amplitude of the SC phase has to locally vanish at a point along the length of the 1D sample in order to have a turn added or removed from it \cite{lit}. Based on this criterion, one looks for an amplitude modulated solution of the SC phase where the amplitude vanishes at least locally (PSC) allowing the SC order parameter to add or remove turns (change its wave number q). Thus, the total phase along the length of the sample can change. An applied voltage across the length of the sample, on one hand, keeps adding turns to the SC order parameter (increases wavenumber) to increase the phase. On the other hand, thermal fluctuations make the system access unstable amplitude modulated states like PSCs to give up phase in the middle of the sample. The LA theory basically rests upon a balance between these two. Although, this vanishing of the amplitude is crucial for such a phenomenon, the LA solution which corresponds to the passage of the system through a free energy saddle never goes to zero anywhere. Interesting to note that, the LA solution can never actually go to zero because of infinite free energy cost it has to incur according to the LA calculations of the free energy and thus it is absolutely against the demand of Little's criterion. LA theory simply claims that by some other fluctuations the solution that comes close to zero would become zero which is wrong in view of the free energy surface as described by LA. Despite this glaring mistake, the LA theory works fine because in reality (on correct free energy surface) the free energy barriers as calculated by LA and the true one (which will be shown) are the same.
\par
In the present paper, we will first clearly identify the error in the identification of the saddle on the free energy surface in LA theory and demonstrate why the good old expression for the variation of amplitude at the bundary of a superconductor under no-field condition holds as an exact solution in the case of LA model as well. Particularly on the basis of this exact solution we will also calculate the corresponding chemical potential profile and the current profile through the sample. We would show that the divergence of the current at the point where the amplitude vanishesh goes against the LA demand of other fluctuations taking a close to zero amplitude actually to zero because, the LA effective free energy can not allow that. we would also argue that the addition or subtraction of turns in the middle of the superconducting sample is not a process that takes place just at the point the amplitude vanishesh, but, should take place rather continuously through the process of formation and relaxation of PSCs.  
\par
The dynamics of a 1D superconductor in the resistive regime is given by time dependent Ginzburg-Landau (TDGL) equation \cite{mac} as 
\ber
(\psi_t+i\mu\psi) &=& \psi_{xx} + (\alpha-\frac{\beta}{2}\vert\psi\vert^2)\psi \\
j&=&Im(\psi^*\bigtriangledown\psi)-\mu_x .
\eer    
In this model, $\psi$ is the superconducting order parameter which is complex valued. The system being 1D, $x$ is distance along the wire from some arbitrary origin. Let us consider, the length of the wire is L with cross sectional area $\sigma$. The $\mu$ is electrochemical potential which can be considered as the order parameter of the N phase within the scope of G-L phenomenology. The constants $\alpha$ and $\beta$ measures the free energy density differences of the SC and N phase as $g_n-g_s=\alpha^2/2\beta=a(\bigtriangleup T^2)$ where $a$ is another constant that depends upon density of states at Fermi surface $N(0)$ and Boltzmann constant as $a=4.7N(0)k_B^2$. The $j$ is current (density) through the sample and the suffix $t$ and $x$ of $\psi$ and $\mu$ indicate of partial derivatives. 
\par
Eq.1 has two stationary solutions. (1) $\psi \equiv 0 $, $\mu = -xj$ (for constant $j$) which is the normal state and (2) $\psi=Ae^{iqx}$, $A^2=(\alpha-q^2)/\beta$, $j=A^2q$, which is the superconducting state when $\mu\equiv 0$ \cite{lan}. The SC order parameter can be visualized as a spiral wound around the x-axis (along the wire). If there are N turns along the length L, there is a total phase difference $\phi = 2\pi N$ along L. Thus, the wave number q of the SC phase is a measure of number of turns the system has on a given length L, since, $q=2\pi N/L$. The expression for the corresponding G-L free energy, when the steady state solution is purely superconducting, is given by 
\bea
F= L\sigma[(q^2-\alpha)A^2+\frac{\beta}{2}A^4]
\eea     
where $\sigma$ is the cross section of the wire. The form of F clearly indicates that, the SC states with smaller q values are energetically favoured. In other words, the spiralling SC order parameter would tend to lose its turns to go to a lower free energy state. Putting the ansatz $\psi=A(x,t)e^{i\phi(x)}$ where $\partial \phi(x)/\partial x = q(x)$ in Eq.1 and separating the real and imaginary parts (where we consider $A(x,t)$ and $\phi(x,t)$ real functions) we get
\ber
\frac{\partial A}{\partial t} - \frac{\partial^2 A}{{\partial x}^2} - (\alpha -q^2)A + \beta A^3=0\\
A\frac{\partial q}{\partial x}+2q\frac{\partial A}{\partial x}-\mu A=0.
\eer
along with these two equations mentioned above one has to take into account the Eq.2 to get a complete picture of the affairs. 
\par
Before we go into the exact solution and calculation of the barriers, let us first have a look at what had gone wrong in the LA calculations of the barrier.The stationary Eq.4 can be rewritten in the form
\bea
\frac{\partial^2 A}{{\partial x}^2} = - \frac{\delta[\frac{(\alpha -q^2)}{2}A^2 - \frac{\beta}{4} A^4]}{\delta A} = -\frac{\delta U}{\delta A}.
\eea
The above expression clearly shows that the effective potential $U$ goes to zero at $A=0$. Whereas, in the LA theory, because of replacing the $q^2A^2/2$ with $-j^2/2A^2$ in the expression of the $U$, it diverges at $A=0$ for all nonzero $j$. This replacement of $q^2A^2/2$ with $j^2/2A^2$ is clearly wrong because the amplitude in the above expression is not the super-current amplitude rather its a modulated form of it and the current density in such a case is not the super-current density corresponding to the constant amplitude SC phase. 
\par
Fig.1a and b show a schematic comparison of the $U$ as taken by LA and us. In our case, the $A=0$ point would be visited by the trajectory from the point $A^2=(\alpha-q^2)/\beta$ where the state is metastable at the cost of an increase of the velocity equivalent $\frac{dA}{dx}$. Interesting to note that, the point $A^2=(\alpha-q^2)/\beta$ is the peak of the $U$ similar to the one considered in LA theory. But, the $U$ of LA does not allow the system to reach the $A=0$ because that would require an infinite storage of kinetic energy equivalent $(\frac{dA}{dx})^2/2$. The actual Ginzburg-Landau free energy containing a gradient square term would then have to keep an infinite free energy storage somewhere in the form of a diverging amplitude gradient in order to allow the system to ever reach the $A=0$ point. But, the constant amplitude SC phase does not allow for such a storage. To circumvent this inconsistency LA proposes this barrier to be a saddle such that the system can escape through other dimensions of the space in which the free energy is defined. A replacement of $q$ by $j$ and $A$ still keeps the space two dimensional and there we see that the $j$ has to fall faster than $A$ and necessarily vanish to keep the free energy divergence free as the $A$ vanishes. This goes against the very concept of current driven origin of the resistive state. Since, the LA calculations mean that the $A=0$ is never accessible for a nonzero $j$, the subsequent claim of LA theory that other fluctuations would make LA solution which goes closer to zero actually reach zero is untenable. As a consequence of a wrong replacement of the wave number the saddle identified is not the lowest barrier and that would be clear in the following where we would show a smaller barrier.  
\begin{figure}
\subfigure []
{\includegraphics[width=4.5 cm,angle=-90]{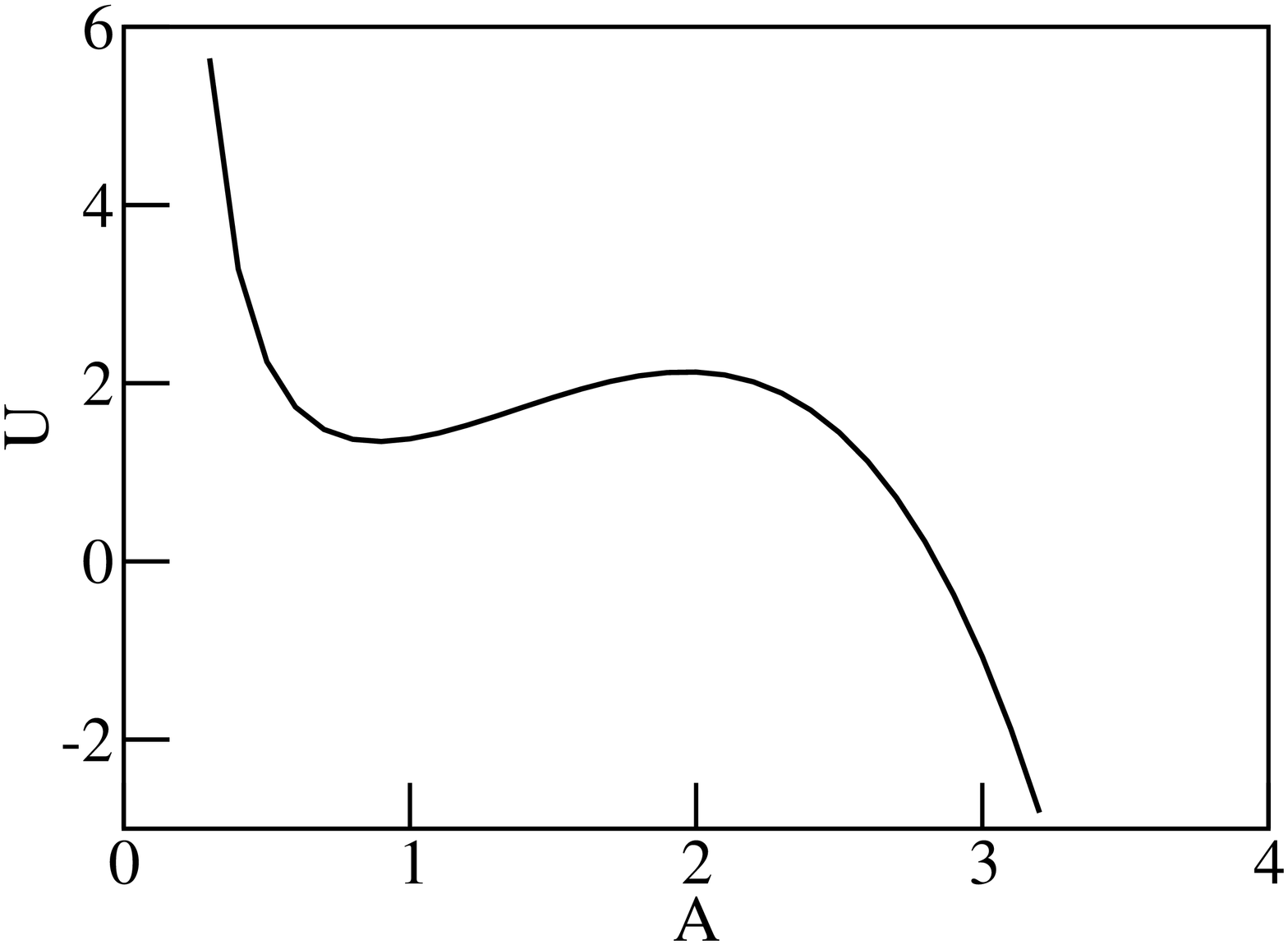}
\label{Fig:a}}
\subfigure []
{\includegraphics[width=4.5 cm,angle=-90]{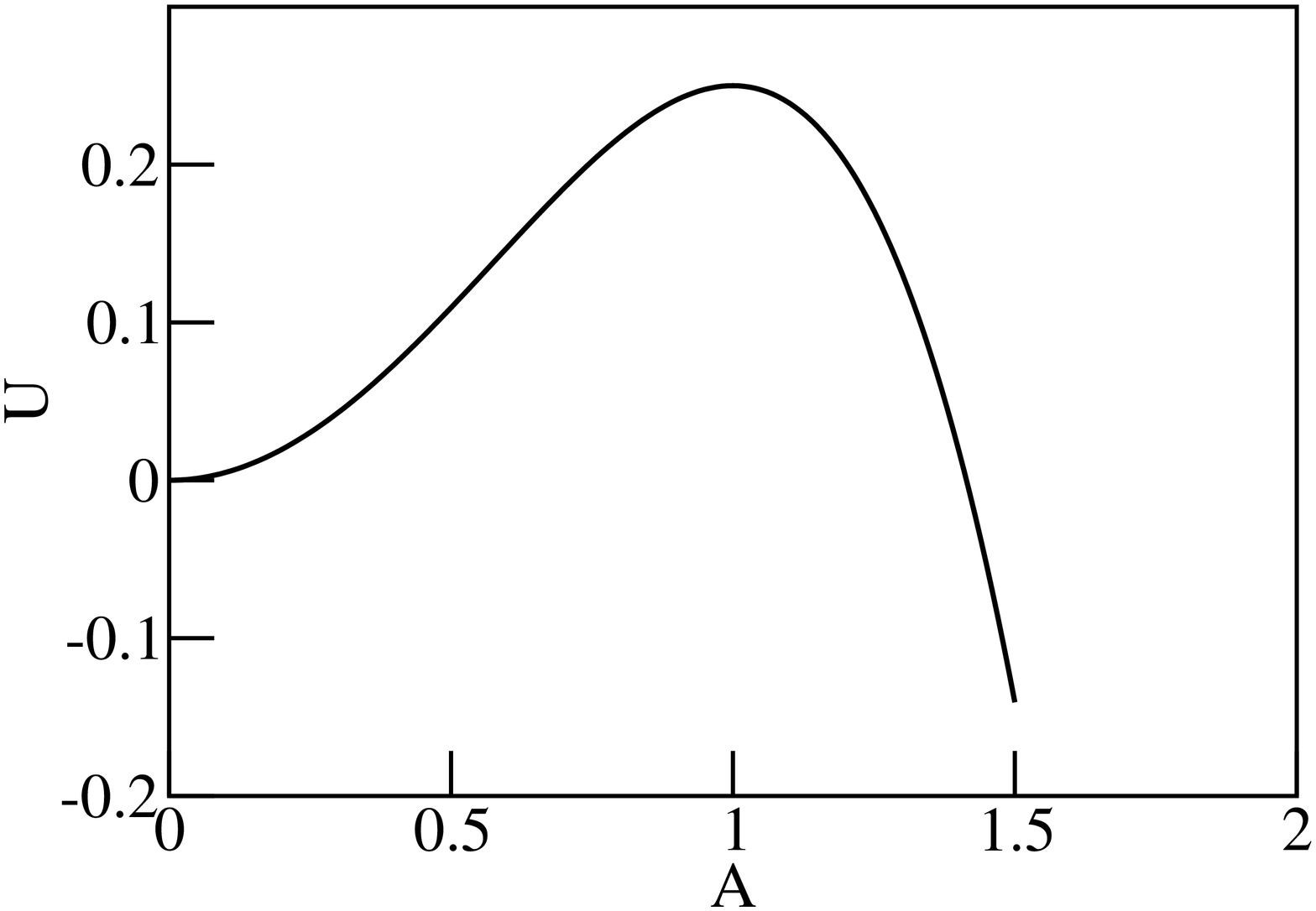}
\label{Fig:b}}
\caption [Figure 1]{\bf schematic diagrams for the comparison of effective potentials on arbitrary scales: (a) $U$ as considered by LA and (b) $U$ that we consider here.}
\end{figure}
\par
Note that, at a constant $q$ the Eq.5 gets a form which had been arrived at by one of the authors \cite{ari1,ari2} previously by a separation of length scales. Eq.5 gives the corresponding $\mu$ profile to an amplitude modulation obtained from Eq.4. In what follows we will stick to this constant $q$ scenario because, a. Eq.4 that actually admits the amplitude modulations is independent of any variation of $q$, b. even with constant $q$ we will be able to show an exact form of the PSC which has a smaller free energy barrier than that proposed by LA and that serves our present purpose and c. this is a good approximation because, normally a single turn gets added or removed at the formation of the PSC and that changes the $q$ by an order $1/L$ which is quite small. Considering the time independence of the amplitude, Eq.4 admits a solution $A=A_0\tanh{x/\sqrt{2}\xi}$ with the couple of conditions - (1) $(\alpha-q^2)-1/\xi^2=0$ and (2) $\beta {A_0}^2 -1/\xi^2=0$ where $\xi$ is the bulk GL coherence length. These conditions immediately identify that ${A_0}^2=(\alpha-q^2)/\beta$ which is the standard amplitude of the steady superconducting state. From the above conditons we recover the standard bulk relationship between the $\alpha$ and $\xi$ also. The above mentioned solution is a well known one which one observes at the boundary a bulk superconductor in no-field conditions. The LA free energy denies it simply because of the divergence of the effective free energy at zero amplitude.
\par
The corresponding $\mu$ profile of the superposed normal phase comes out to be $\mu=(2q/\sqrt{2}\xi)(1/\tanh{x/\sqrt{2}\xi}-\tanh{x/\sqrt{2}\xi})$. This is the shouted after exact solution which would corresponds to an energy barrier same as that in LA theory. The barrier height would become even smaller with a nonzero $q$ as is expected from GL free energy. Corresponding to our exact solutions for $A$ and $\mu$ the current can be evaluated from Eq.2 for a constant $q$. Its important to note that, this current has a value $A_0^2q$ far from the origin and it diverges at the origin as $1/\tanh^2{x/\sqrt{2}\xi}$. So, it has a part which is grows as $1/A^2$ where $A\to 0$ and there is no question of the current falling faster than $A$ for the divergence in the effective free energy $U$ of LA to remain analytic. So, according to the free energy expression of the LA this exact solution of the system does not qualify for overcoming the free energy barrier but so far done experimental studies basically confirm having such solutions that vanish locally.
\par
At the limit the super-current density $j\to 0$ the very complicated looking LA solution through the LA mentioned saddle of the free energy surface actually overcomes a free energy barrier somewhat bigger than $\bigtriangleup F=(8\sqrt{2}/3K_BT)(g_n-g_s)\sigma\xi$ \cite{lan,tink,kop}. The free energy barrier corresponding to our exact PSC solution $A=A_0\tanh{x/\sqrt{2}\xi}$ can be calculated by putting $\psi=A_0\tanh{x/\sqrt{2}\xi}e^{iqx}$ and $A_0e^{iqx}$ respectively in the expression of the free energy (corresponding to Eq.1) shown below and subtracting the latter free energy from the former
\bea
F = \sigma \int_{-\infty}^{\infty}{dx[|\bigtriangledown \psi|^2-\alpha|\psi|^2 + \frac{\beta}{2}|\psi|^4]}.
\eea  
If we make use of the conditions (1) and (2) as mentioned above in connection with the derivation of the PSC we can immediately show that the free energy barrier is exactly calculated as 
\bea
\bigtriangleup F=(8\sqrt{2}/3K_BT)(g_n-g_s)\sigma\xi - \frac{26\sqrt{2}\sigma}{3\beta\xi}q^2 + \frac{28\sigma\xi}{3\sqrt{2}\beta}q^4.
\eea
We can see that the $q=0$ that corresponds well with the $j=A^2q=0$ condition, the free energy barrier is clearly the same as the LA value when the relevant length scale of the amplitude modulation is equal to the bulk GL coherence length $\xi$. Taking roughly $q \to 1/\xi$ one recovers the $\bigtriangleup T^{2/3}$ power law in all the terms of the above mentioned free energy expression. Note that, consideration of a nonzero $q$ even reduces the free energy barrier. Important to note that $q=0$, actually sets $\mu=0$ which is a violation of the relation 
\bea
\frac{4\pi e}{h}\bigtriangleup V = \frac{\partial \bigtriangleup arg\psi}{\partial t},
\eea
where $\bigtriangleup V$ is the applied voltage drop across the system. In literature, the limit at which the $\bigtriangleup T$ becomes $\bigtriangleup T_C$ is $j/j_C\to 0$ where $j_C=2\alpha^{3/2}/3\sqrt{3}\beta$. This limit can be seen as $q\to 0$ limit also, but, $q$ cannot actually vanish for the theory to be consistent and so is $j$. Taking into account that, a nonzero $q$ is essential one can easily infer that a nonzero $A$ and $j$, however small, should accompany a PSC and that the LA solution of the PSC would actually encounter an enormous barrier. 
\par
Let us have a discussion on the implication of the present results. The LAMH theory is the most useful and accepted framework in understanding classical PSC induced origin of the resistive regime of the 1D superconductor. It matches pretty well the experimentally obtained resistivity vs temperature plots. Our present analysis helps solve the riddle as to how such a successful theory can actually not predict a proper PSC solution and establishes LAMH theory on even stronger foundation after more than 40 years of its introduction. We clearly show that, it was the wrong identification of the free energy barrier which resulted in this discrepancy in the value of the $\bigtriangleup T_C$ by finding out an exact solution of the same dynamical equation used by LAMH and corresponding free energy barrier. The LA theory, however, uses the correct numerical value despite identifying the wrong barrier because it had stuck to the conservation of energy in getting the saddle. Here, we actually show that no such saddle exists because the current density can not go to zero at the point as the amplitude does. We have shown that the good old solution for the amplitude modulations at the boundary of a bulk superconductor holds perfectly good in the LA case as well.
\par
Our present analysis apart from resolving this PSC riddle also clarifies and sheds light to a few other points. First of all, we have considered here a constant $q$ during the PSC formation. This not only simplifies the calculations but comes out to be correct because, in general only a turn addition or removal happens by the formation of a PSC. Moreover, as we have already shown that, there is no reason to think of a coupling between the amplitude and the wave number in the modulated amplitude phase just as in the SC phase and that has severe consequences. It has to be explored what other forms of PSC like or other solutions exist with a spatially varying $q$. Nevertheless, in the present context, with a constant $q$ we are able to point out the discrepancy in the calculations of the saddle in LA theory. Our solution which satisfies the Little's criterion of amplitude vanishing is clearly having the same barrier to overcome as LA's saddle and in that sense is a better representative for PSC. Note that, our exact solution is unstable for $\tanh^2{(x/\sqrt{2}\xi)}<1/3$ to any infinitesimal uniform perturbation. So, fluctuations would start growing at the core of the PSC and would make it relax back to the constant amplitude SC phase.
\par
Another important point to note is that, the claim that the LA solution approximates to a form proportional to $\tanh{|x|}$ \cite{kop} is somewhat in conflict with the Eq.4 because, this form would produce a delta function contribution at the origin in Eq.4 resulting in an invalidity of such a form as a solution. This is why our exact solution is of the form $\tanh{(x/\sqrt{2}\xi)}$ which induces a global phase shift of $\pi$ on one half of it unlike the $\tanh{|x|}$ form. Its interesting to note that, the Eq.1 is invariant under a constant global phase shift. As a result, such a phase shift should not be energetically unfavourable and can happen with a locally vanishing amplitude particularly because the single valued nature of the SC order parameter is not compromised. But, it marks a little departure from the Little's proposition that a phase change of $2\pi$ or its multiple would actually happen right at the vanishing of the amplitude. Rather, we see here that a phase change of half of that required to add/lose a turn is happening at the time of the formation of the PSC and the other half has to happen during the relaxation of the PSC to keep the order parameter single valued. One way of looking at this scenario would be that, as the PSC forms, it produces strain locally to the turns and that strain relaxes by rotating one half of the spiral. By the time the PSC has occurred, half of the spiral is rotated by an angle $\pi$. Now a relaxation of the PSC might find it energetically favourable or retaines some memory to continue effectively rotating the same half in the same direction so that when the relaxation of the PSC is complete the system loses/gains a turn. A detailed investigation using the dynamics could be revealing. We conclude by saying that, we have exactly solved the TDGL model for the resistive state of an one dimensional superconductor to explicitly show a similar energy barrier as the LA saddle. On the basis of that we identify the error in the LA calculation of energy barrier. We put the LAMH theory on a even firmer basis.

\end{document}